\documentclass{article}

\usepackage{graphicx}

\def\a{\alpha}

\def\ga{\gamma}
\def\s{\sigma}

\def\^#1{\widehat{#1}}

\def\beql#1{\begin{equation} \label{#1}}
\def\beq{\begin{equation}}
\def\eeq{\end{equation}}

\def\<{\langle}
\def\>{\rangle}

\def\eqref#1{(\ref{#1})}

\begin{document}

\title{Data Analysis for the COVID-19 \\ early dynamics in Northern Italy. \\ The effect of first restrictive measures}

\author{G. Gaeta \\ {\it Dipartimento di Matematica, Universita' degli Studi di Milano} \\ {\it via Saldini 50, I-20133 Milano (Italy)} \\ and \\ {\it SMRI, 00058 Santa Marinella (Italy)} \\ {\tt giuseppe.gaeta@unimi.it}
}

\maketitle

\begin{abstract}

In a recent report we have collected some data about the COVID-19 epidemics in Northern Italy; in this follow-up we analyze how these changed after the mild restrictive measures taken by the Government two weeks ago and the large campaign of public awareness developed in the meanwhile.

\end{abstract}

In a recent report \cite{Gcov} we have collected some data about the COVID-19 epidemics in Northern Italy; in particular, we have analyzed and fitted them by an exponential law in order to extract the growth factor both at nationwide level and at the local one. The purpose of this follow-up is to analyze how these changed after the mild restrictive measures taken by the Government and the campaign of public awareness.

These started at February 24; the effect of such measures is of course showing up with some delay, corresponding to the incubation time of COVID-19.

The outcome of our analysis is that there was a slowing down of the epidemics, but this is still too weak to face the menace of a large scale epidemics. Actually, more stringent restrictions went on operation from March 8, and our analysis shows that these were fully justified.

Our analysis is purely at the statistical level over available data; that is, we will not discuss any model nor try to interpret the data in view of several theories circulating in the scientific communities; in particular we do not try to estimate the real number of infections, which according to certain analysis could be from two to three times the number of known cases.

The data for countries other than Italy are extracted from the ``situation reports'' of the World Health Organization \cite{WHOrep}; those for Italy are extracted from ``Ministero della Salute'' and ``Protezione Civile'' (a governmental agency) \cite{MinSal,PCrep}.

In all cases, we consider -- as predicted by virtually all epidemiological models for the initial phase of an epidemic \cite{Murray,Edel,Heth} -- an exponential law for the number of infected people,
\beql{eq:alpha} n(t) \ = \ \exp[ \a \, t ] \ n(0) \eeq
and tried to fit $\a$ the growth exponent $\a$ from available data. (We will always use one day as the time unit.)

Two other relevant epidemiological parameters are simply related to $\a$.
The doubling time $\tau$ is the time needed to double $n(t)$, i.e. such that $n(t + \tau) = 2 n(t)$; this is obtained from the above via
\beql{eq:tau} \tau \ = \ \a^{-1} \ \log (2) \ . \eeq
The daily growth factor $\gamma$,such that $n(t+1) = \gamma n(t)$, is determined as
\beql{eq:gamma} \ga \ = \ \exp [ \a ] \ . \eeq

\section{Benchmarks: China, Korea}

In \cite{Gcov} we have analyzed data from China and Korea in order to have a term of comparison. Tables reporting the (updated to March 7) data for these countries are reported in Appendix A. We are interested in fitting these data with an exponential law, considering limited timespans; in particular we are interested in considering how the restrictive measures adopted by the Chinese and Korean Governments influenced the growth factor $\a$.

In order to do this we considered in both cases an ``initial'' and a ``final'' (or actually a ``recent'') timespan; in order to make a more direct comparison, we decided here -- at difference with what was done in \cite{Gcov} -- to consider in all cases a period of one week.

In the case of China the initial period was that of January 23 to February 2, the final one that of February 27 to March 7; while for Republic of Korea the initial period it was that of February 18 to February 24, and the final one from March 1 to March 7. We denote by a subscript ``$i$'' and by a subscript ``$f$'' the quantities referring to the initial and final week respectively. We will also consider the simplest measure of the reduction of the epidemic speed, i.e.  the ratio $r = \a_f/\a_i$.

\begin{figure}
\begin{tabular}{cc}
\includegraphics[width=150pt]{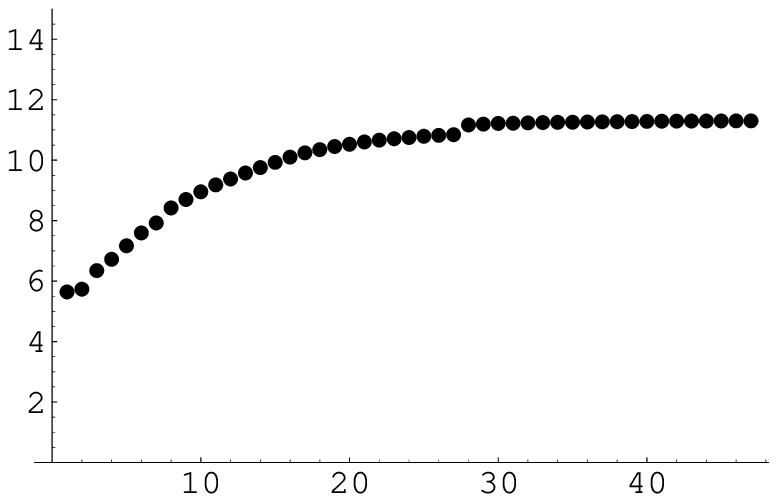} &
\includegraphics[width=150pt]{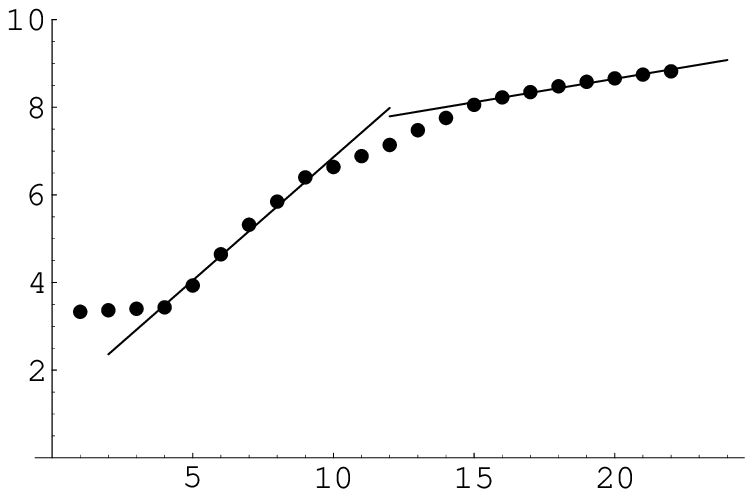} \\
 China & Korea \end{tabular}
  \caption{Semi-logarithmic plots for the data of China and Korea.}\label{fig:ChinaKorea}
\end{figure}

The result of the analysis for China and Korea is summarized in Table I.a below. See also Figure \ref{fig:ChinaKorea}.

\begin{center}
\begin{tabular}{||l||l|l||}
\hline
 & China & Korea \\
\hline
$\a_i$   & 0.33 & 0.56 \\
$\tau_i$ & 2.10 & 1.23 \\
$\ga_i$  & 1.39 & 1.75 \\
\hline
$\a_f$   & 0.003 & 0.11 \\
$\tau_f$ & 212 & 6.47 \\
$\ga_f$  & 1.003 & 1.11 \\
\hline
$r$ & 0.009 & 0.20 \\
\hline
\end{tabular}

\medskip
{\tt Table I.a.} Epidemiological parameters for China and Korea in the initial phase and in the last week, i.e. after the restrictive measures. See text.
\end{center}
\bigskip

\section{Other European countries: France, \\ Germany and Spain}

\begin{figure}
  \begin{tabular}{ccc}
  \includegraphics[width=100pt]{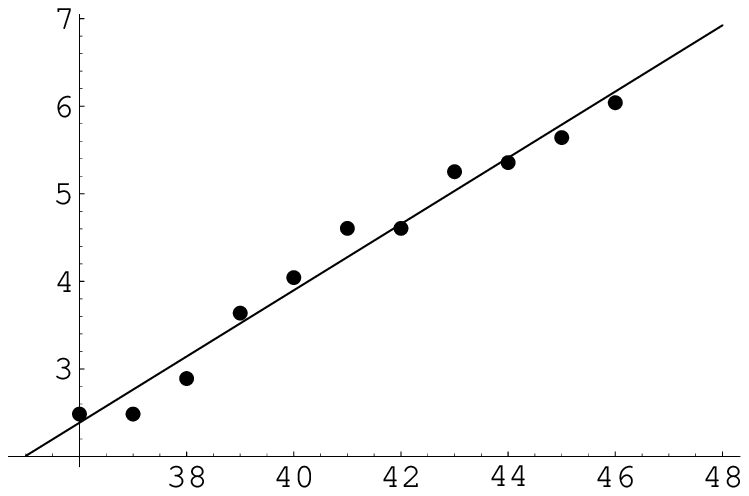} &
  \includegraphics[width=100pt]{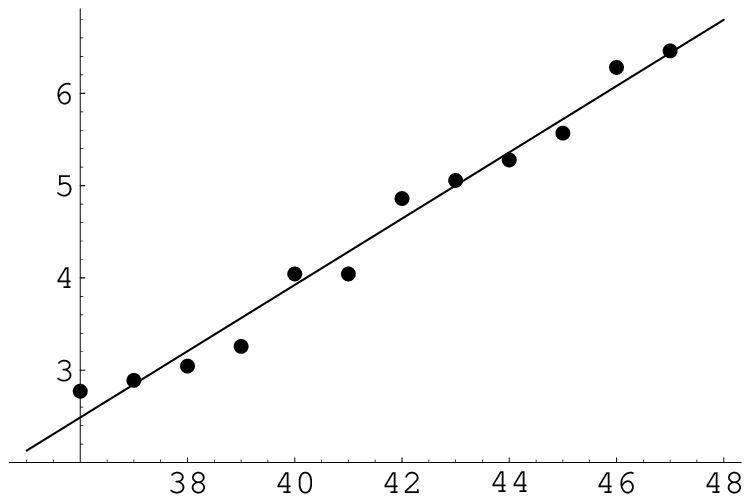} &
  \includegraphics[width=100pt]{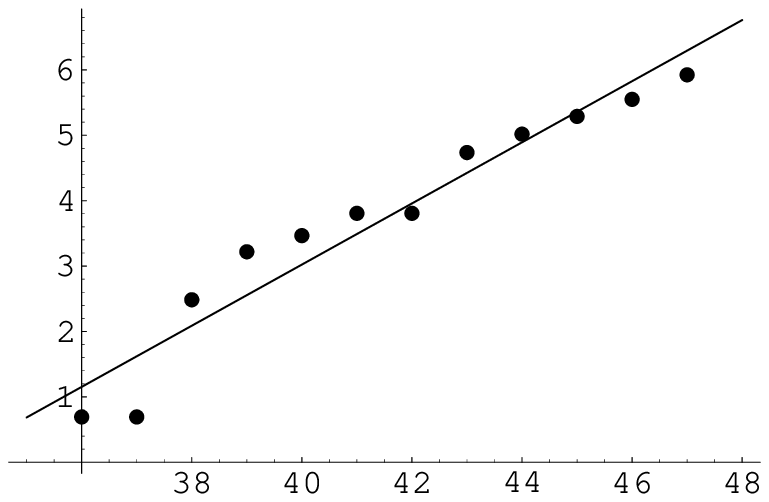} \\
  France & Germany & Spain \end{tabular}
  \caption{Semi-logarithmic plots for the data of different European countries.}\label{fig:Euro}
 \end{figure}

In the last few days the COVID-19 epidemics reached other continental European countries; the data for some of these -- in particular, France, Germany and Spain -- are also given in Appendix A. In this case there were no substantial restrictive measures yet,nor any time to observe a change in the trend, so we are only able to make a measurement of the growth exponent in the initial phase of the epidemics and to compare it with the one observed in Italy. The parameters for France, Germany and Spain are given in Table I.b below; see also Figure \ref{fig:Euro}.

\begin{center}
\begin{tabular}{||l||c|c|c||}
\hline
 & F & D & E \\
\hline
$\a$   & 0.38 & 0.36 & 0.47 \\
$\tau$ & 1.83 & 1.93 & 1.48 \\
$\ga$  & 1.46 & 1.43 & 1.60 \\
\hline
\end{tabular}

\medskip
{\tt Table I.b.} Epidemiological parameters for France (F), Germany (D) and Spain (E); all of them are in the initial epidemic phase. See text.
\end{center}
\bigskip

\section{Italy: national and regional data}

The data for Italy (I) and for the regions more heavily touched by the virus, i.e. Lombardia (L), Veneto (V) and Emilia-Romagna (ER) are reported in Appendix B. These time series are shorter than the ones available for China and Korea, and for these day one is February 21; moreover the first three days show a very steep increase of cases, which could be due to late recognition of infections present since some time. Also, restrictive measures were taken on February 24 and in view of the incubation time of COVID-19 they could show their effect only after about ten days. We decided therefore to consider as initial period the days from February 24 to February 28 (inclusive), and as final period the days from March 3 to March 7 (inclusive).

We measure the achieved reduction in the speed of the epidemic by the parameter
\beql{eq:r} r \ := \ \frac{\a_f}{\a_i} \ . \eeq

\begin{figure}
\begin{tabular}{|c|c|}
\hline
\includegraphics[width=150pt]{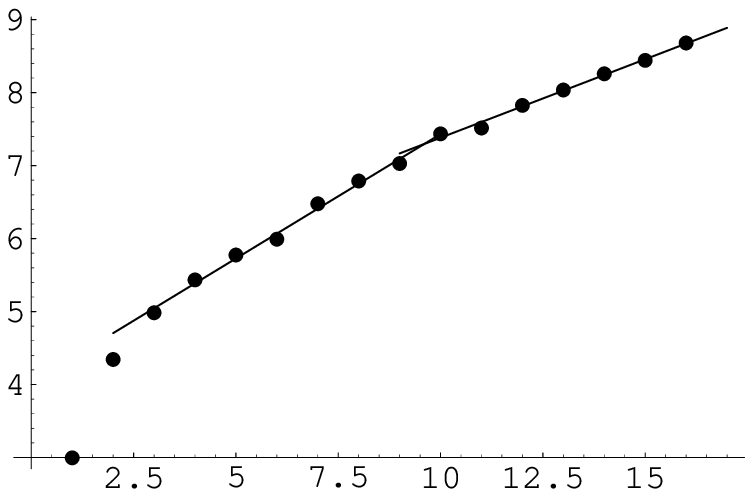} &
\includegraphics[width=150pt]{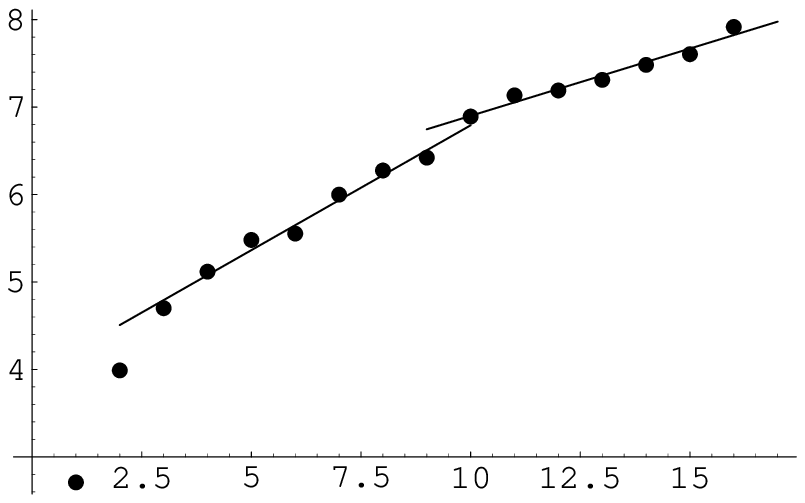} \\
Italy & Lombardia \\
\hline
\includegraphics[width=150pt]{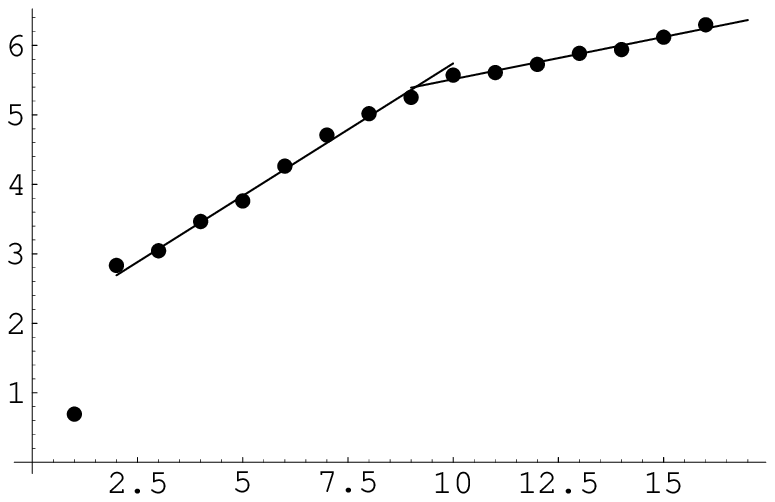} &
\includegraphics[width=150pt]{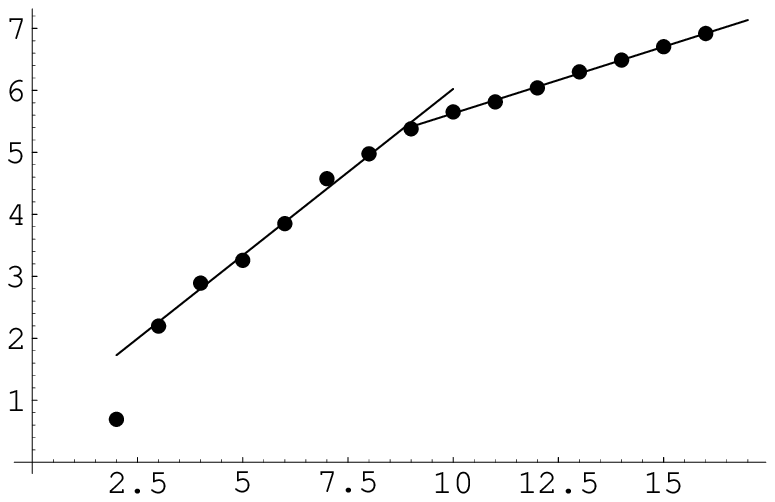} \\
Veneto & Emilia-Romagna \\
\hline
\end{tabular}
  \caption{Semi-logarithmic plots and exponential fits in the initial and final period for the whole of Italy and for specific regions. See text.}\label{fig:Ita}
\end{figure}

The results of the analysis are summarized in Table II; see also Figure \ref{fig:Ita}. The latter shows that the exponential fit is in all considered cases quite good.

\begin{center}
\begin{tabular}{||l||c||c|c|c||}
\hline
         & I & L & V & ER \\
\hline
$\a_i$   & 0.34 & 0.29 & 0.38 & 0.54 \\
$\tau_i$ & 2.03 & 2.43 & 1.82 & 1.29 \\
$\ga_i$  & 1.41 & 1.33 & 1.46 & 1.71 \\
\hline
$\a_f$   & 0.22 & 0.15 & 0.12 & 0.22 \\
$\tau_f$ & 3.22 & 4.50 & 5.70 & 3.22 \\
$\ga_f$  & 1.24 & 1.17 & 1.13 & 1.24 \\
\hline
$r$      & 0.88 & 0.88 & 0.77 & 0.72 \\
\hline
\end{tabular}

\medskip
{\tt Table II.} Epidemiological parameters for Italy (I) and for the regions of Lombardia (L), Veneto (V) and Emilia-Romagna (ER), in the initial phase and in the last week, i.e. after the restrictive measures. The parameter $r := \a_f / \a_i$ is a measure of the achieved reduction in the epidemic speed. See text.
\end{center}
\bigskip

\bigskip

It is interesting to consider two other parameters, i.e. the ratio $\rho$ of infected people in Home isolation -- thus presumably with light or no symptoms -- to the total of infective people (note this includes only those for which the infection is active), and the ratio $\s$ of patients in Intensive Care units to the total of hospitalized patients. The latter is possibly confused in the last days, as ``sub-intensive care'' units were created. As for the former, this is possibly influenced by the fact that in the presence of a large number of infected with heavy symptoms there is little chance to analyze people with light symptoms or who just had contacts with known infected but show no symptoms; this could explain why $\rho$ remained around $0.5$ for several days to fall down as the number of infections raised sharply. Data for the quantities $\rho$ and $\s$ are also given in Appendix B.

\section{Local communities in Northern Italy}

In our previous study, we had considered a few Departments in Northern Italy, mostly located near the area which was more heavily struck by COVID-19.

Data for these -- i.e. Lodi (LO), Cremona (CR), Piacenza (PC), Pavia (PV), Bergamo (BG), Brescia (BS), Milano (MI) and Padova (PD) -- are given in Appendix C; here we also provide data for other Departments which were not considered in \cite{Gcov} but do now show some worrying evolution, i.e. Venezia (VE), Treviso (TV), Rimini (RN) and Pesaro (PU).

The initial and final period considered for these Departments are the same considered for the nationwide and regional analysis in Italy, i.e. February 24 to February 28 (inclusive) and March 3 to March 7 (inclusive).

\begin{figure}
\begin{tabular}{|c|c|}
\hline
  \includegraphics[width=150pt]{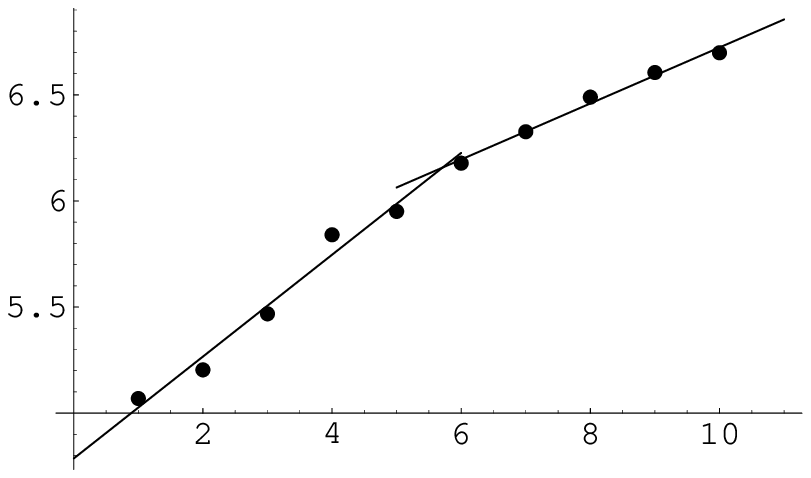} &
  \includegraphics[width=150pt]{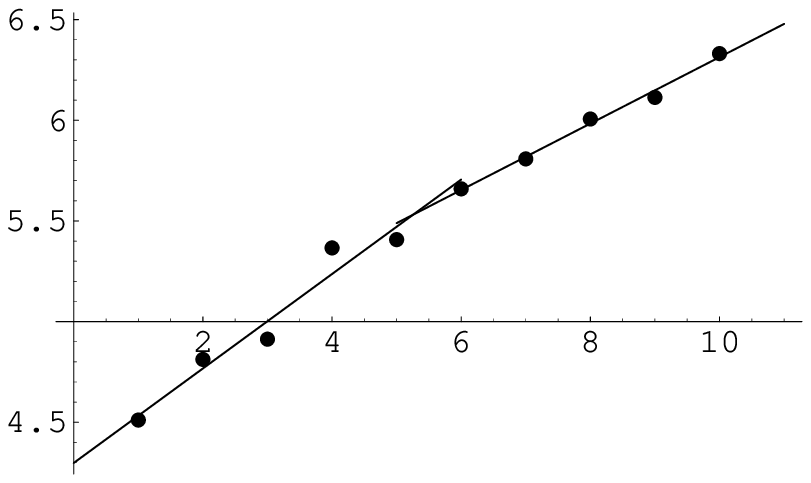} \\
  LO & CR \\
  \hline
  \includegraphics[width=150pt]{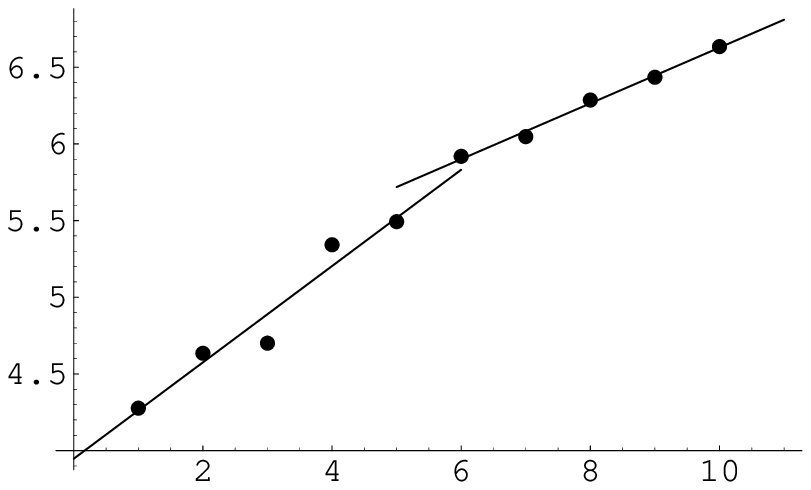} &
  \includegraphics[width=150pt]{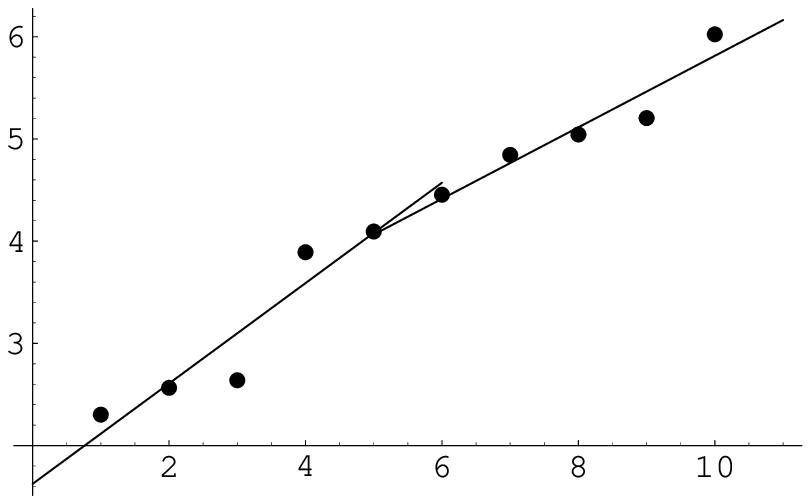} \\
  BG & BS \\
  \hline
  \includegraphics[width=150pt]{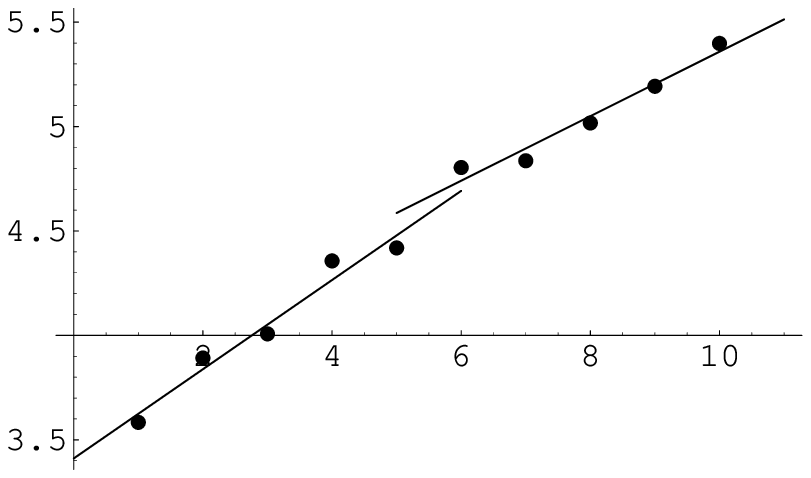} &
  \includegraphics[width=150pt]{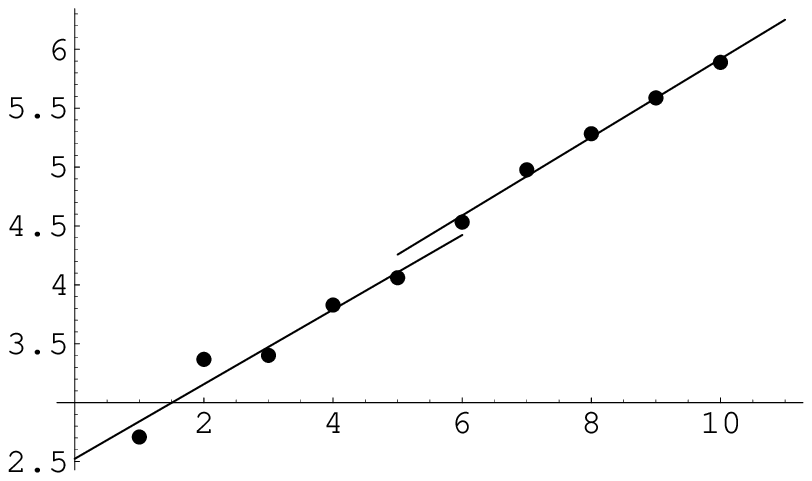} \\
  PV & MI \\
\hline
\end{tabular}
  \caption{Semi-logarithmic plots and exponential fits in the initial and most recent phase of COVID epidemics in different considered Departments within the region of Lombardia. See text}\label{fig:prov1}
\end{figure}

\begin{figure}
\begin{tabular}{|c|c|}
\hline
  \includegraphics[width=150pt]{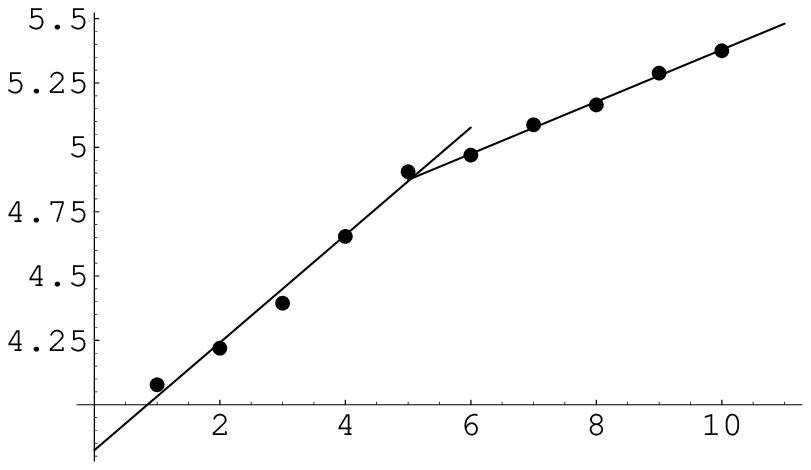} &
  \includegraphics[width=150pt]{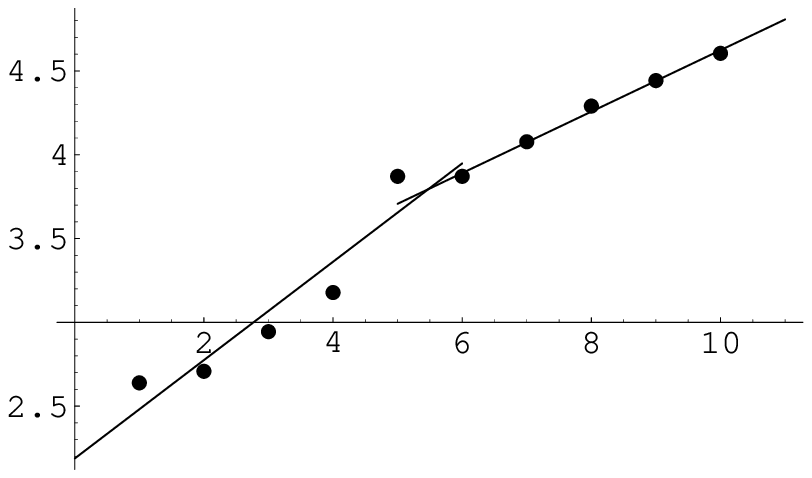} \\
  PD & VE \\
  \hline
  \includegraphics[width=150pt]{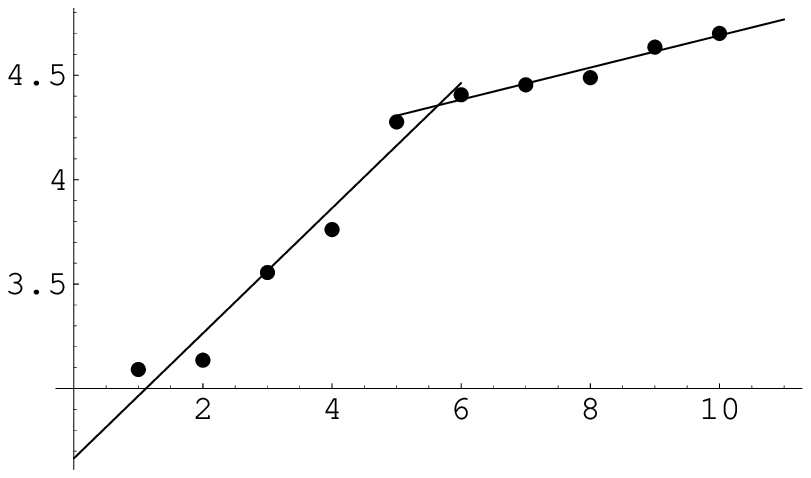} &
  \includegraphics[width=150pt]{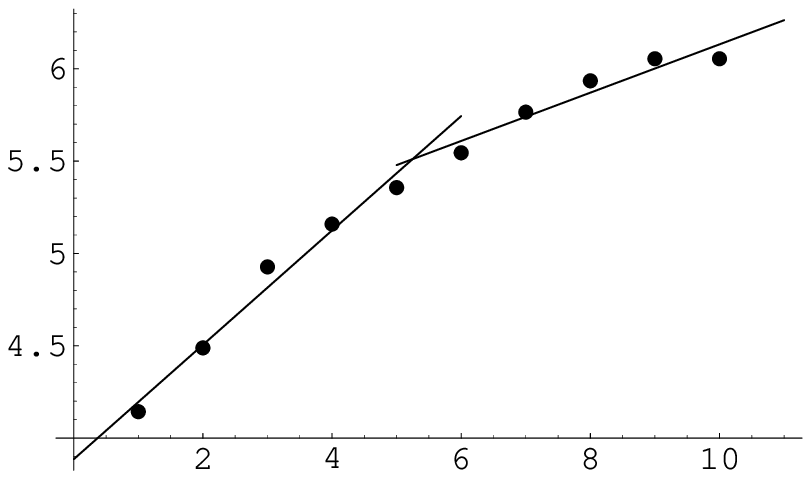} \\
  TV & PC \\
  \hline
  \includegraphics[width=150pt]{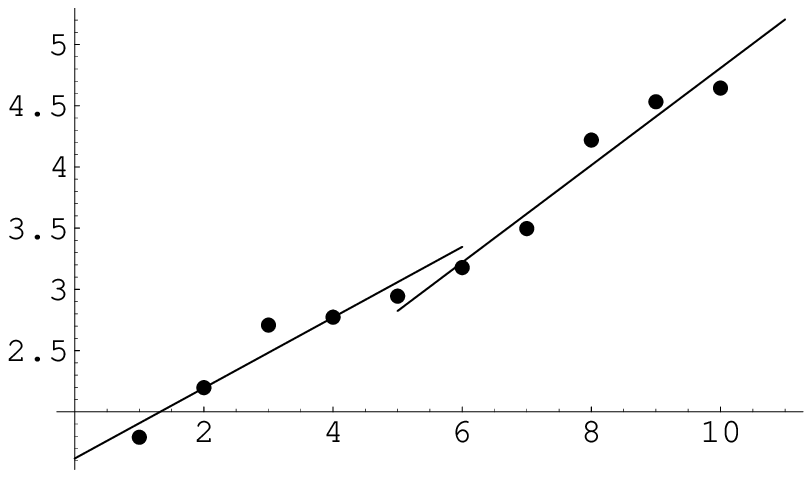} &
  \includegraphics[width=150pt]{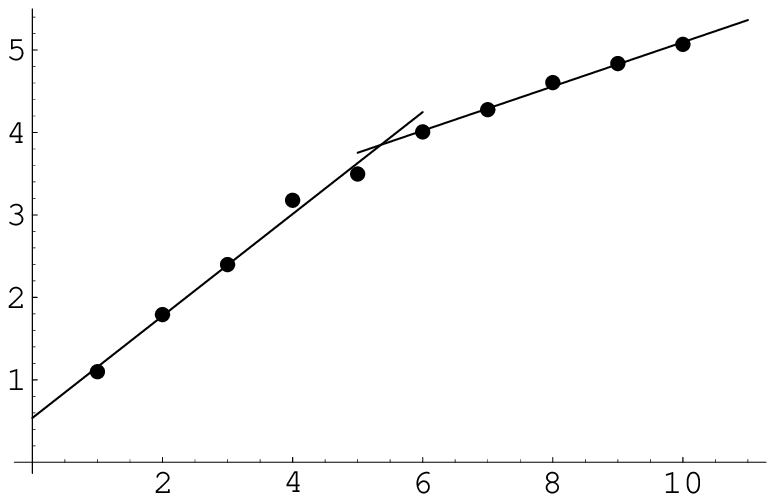} \\
  RN & PU \\
\hline
\end{tabular}
  \caption{Semi-logarithmic plots and exponential fits in the initial and most recent phase of COVID epidemics in different considered Departments within the region of Veneto (PD, VE, TV), Emilia-Romagna (PC,RN) and Marche (PU). See text}\label{fig:prov2}
\end{figure}

The results of the analysis are summarized in Table III; see also Figures \ref{fig:prov1} and \ref{fig:prov2}. Note that while the exponential fits are usually quite good, in some cases -- where large fluctuations in the new data, possibly due to delay in registration of new infections, seem to be present -- they appear to be less reliable.

\begin{center}
{\small
\begin{tabular}{|l||r|r|r|r||r|r|r|r||r|r|r|r||}
\hline
       &  LO  &  CR  &  PC  &  PV  &  BG  &  BS  &  MI  &  PD  &  VE  &  TV  &  RN &  PU \\
\hline
$\a_i$   & 0.24 & 0.23 & 0.31 & 0.21 & 0.31 & 0.49 & 0.32 &
0.2 1 &  0.29 & 0.30 & 0.29 & 0.62 \\
$\tau_i$ & 2.89 & 2.95 & 2.24 & 3.25 & 2.21 & 1.41 & 2.19 & 3.31 &
2.36 & 2.31 & 2.41 & 1.12 \\
$\ga_i$  & 1.27 & 1.26 & 1.36 & 1.24 & 1.37 & 1.63 & 1.37 & 1.23 &
1.34 & 1.35 & 1.33 & 1.86 \\
\hline
$\a_f$   & 0.13 & 0.16 & 0.13 & 0.15 & 0.18 & 0.35 & 0.33 &
0.10 & 0.18 & 0.08 & 0.40 & 0.27 \\
$\tau_f$ & 5.25 & 4.20 & 5.30 & 4.49 & 3.81 & 1.98 & 2.09 & 6.85 &
3.78 & 9.03& 1.75 & 2.58  \\
$\ga_f$  & 1.14 & 1.18 & 1.14 & 1.17 & 1.20 & 1.42 & 1.39 & 1.11 &
1.20 & 1.08 & 1.49& 1.31 \\
\hline
r & 0.55 & 0.70 & 0.42 & 0.72 & 0.58 & 0.71 & 1.05 &
0.48 & 0.62 & 0.26 & 1.38 & 0.43 \\
\hline
\end{tabular}
}

\medskip

{\tt Table III.} Best fit of the $\a$ factor with the corresponding doubling time $\tau$ and daily growth factor $\ga$ -- see eqs. \eqref{eq:alpha}, \eqref{eq:tau} and \eqref{eq:gamma} -- for the different Northern Italy Departments considered. The reduction parameter $r$, see \eqref{eq:r}, is also computed. See Figures \ref{fig:prov1} and \ref{fig:prov2} for the fit.
\bigskip
\end{center}

\section{Discussion and conclusions}

We have analyzed how the epidemiological data, in particular the exponential growth rate $\a$ defined by \eqref{eq:alpha} and the related quantities $\tau$ and $\ga$, see \eqref{eq:tau} and \eqref{eq:gamma}, changed from the initial phase of the epidemics in Northern Italy to the last days. In particular, this allowed to give an evaluation of the effect of the mild restrictive measures taken by the Italian Government and of the large-scale public awareness campaign going on in the country.

The main parameter to evaluate this effect is the reduction factor $r$ defined in \eqref{eq:r}. In the case of the Republic of Korea, i.e. a country with a similar total population and political system but different restrictive measures, this turned out to be $r = 0.20$.

In the case of Italy, we estimated $r = 0.88$; the reduction factor is however quite different in different regions and -- as had to be expected on statistical basis -- even more so when we look at a finer scale, i.e. for different Departments, see Table III. It turns out from this table that in some Department (in particular, among those considered here, Milano and Rimini -- note however that for the latter the exponential fit is not very good) the epidemic speed has indeed grown up in the last period. In other Departments the reduction has been from moderate, around $r=0.7$ to discrete, i.e. around $r = 0.45$, and in one case (Treviso) quite good, with $r=0.26$, and comparable with that of Korea  despite the restrictive measures taken here were much milder than in Korea.

In any case, the reduction achieved is encouraging, but clearly far from being sufficient to stop or even substantially slow down the epidemic dynamics\footnote{See also the discussion in \cite{Gcov} in this respect.}; a much more substantial reduction is required for this.

We have also considered the initial epidemic dynamics in other European countries,in particular France, Germany and Spain; here the data analyzed concerned the whole of the countries. We noted that the exponential growth rate in these countries range from 0.36 of Germany to 0.47 of Spain, with an intermediate 0.38 for France. These worrying figures should be compared with the initial growth rate in Italy, i.e. 0.34 for the whole country with 0.29 for the most affected region, i.e. Lombardia; they mean the possibility of a large COVID-19 epidemic in these countries in the next weeks should not be discarded unless prompt action is taken there too.

\section*{Acknowledgements}

I thank L. Peliti, M. Cadoni and E. Franco for discussions; the opinions expressed in this report are entirely of the author and do not entail any responsibility of other individuals. The paper was prepared over a stay at SMRI. The author is also a member of GNFM-INdAM.

\newpage

\begin{appendix}

\section{Data for foreign countries}

\subsection{China and Korea}

\begin{center}

{\small

\begin{tabular}{|r|r||r|r||r|r||r|r||r|r||}
\hline
day & cases & day & cases & day & cases & day & cases & day & cases \\
\hline
1 &  282 & 2 & 309 & 3 & 571 & 4 & 830 & 5 & 1297 \\
6 & 1985 & 7 & 2761 & 8 & 4537 & 9 & 5997 & 10 & 7736 \\
11 & 9720 & 12 & 11821 & 13 & 14411 & 14 & 17238 & 15 & 20471 \\
16 & 24363 & 17 & 28060 & 18 & 31211 & 19 & 34598 & 20 & 37251 \\
21 & 40235 & 22 & 42708 & 23 & 44730 & 24 & 46550 & 25 & 48548 \\
26 & 50054 & 27 & 51174 & 28 & 70635 & 29 & 72528 & 30 & 74280 \\
31 & 74675 & 32 & 75569 & 33 & 76392 & 34 & 77042 & 35 & 77262 \\
36 & 77780 & 37 & 78191 & 38 & 78630 & 39 & 78961 & 40 & 79394 \\
41 & 79968 & 42 & 80174 & 43 & 80304 & 44 & 80422 & 45 & 80565 \\
46 & 80711 & 47 & 80813 &    &       &    &       &    &       \\
\hline
\end{tabular}

}

\medskip

{\tt Table A.I.} COVID-19 cases in China; day 1 is January 21. Source: WHO situation reports \cite{WHOrep}. Note that on Day 28 (February 17) the method of counting was changed (clinical evidence being considered sufficient even without laboratory test), leading to a sudden jump in the number of cases.

\bigskip
    \end{center}

\begin{center}
{\small

\begin{tabular}{|r|r||r|r||r|r||r|r||r|r||}
\hline
day & cases & day & cases & day & cases & day & cases & day & cases \\
\hline
 1 &   30 &  2 &   31 &  3 &   51 &  4 &  104 &  5 &  204 \\
 6 &  346 &  7 &  602 &  8 &  763 &  9 &  977 & 10 & 1261 \\
11 & 1766 & 12 & 2337 & 13 & 3150 & 14 & 3736 & 15 & 4212 \\
16 & 4812 & 17 & 5328 & 18 & 5766 & 19 & 6284 & 20 & 6767 \\
\hline
\end{tabular}

}

\medskip

{\tt Table A.II.} COVID-19 cases in the Republic of Korea; day 1 is February 17. Source: WHO situation reports \cite{WHOrep}.

\bigskip
    \end{center}

\subsection{Western Europe countries}

\begin{center}

\begin{tabular}{|l||r|r|r||}
\hline
day & F & D & E \\
\hline
25 Feb &   12 &  16 &  2  \\
26 Feb &   12 &  18 &  2  \\
27 Feb &   18 &  21 & 12  \\
28 Feb &   38 &  26 & 25  \\
29 Feb &   57 &  57 & 32  \\
01 Mar &  100 &  57 & 45  \\
02 Mar &  100 & 129 & 45  \\
03 Mar &  191 & 157 & 114  \\
04 Mar &  212 & 196 & 151  \\
05 Mar &  282 & 262 & 198  \\
06 Mar &  420 & 534 & 257  \\
07 Mar &  613 & 639 & 374  \\
\hline
\end{tabular}

\medskip

{\tt Table A.III.} Known cases of contagion in France (F), Germany (D) and Spain (S)  \cite{WHOrep}.
\end{center}
\bigskip

\section{Italy. Nationwide and regional data}

\begin{center}

\begin{tabular}{|l||r||r|r|r||}
\hline
day & I & L & V & ER \\
\hline
21 Feb &   20 &   15 &   2 &   0 \\
22 Feb &   77 &   54 &  17 &   2 \\
23 Feb &  146 &  110 &  21 &   9 \\
24 Feb &  229 &  167 &  32 &  18 \\
25 Feb &  322 &  240 &  43 &  26 \\
26 Feb &  400 &  258 &  71 &  47 \\
27 Feb &  650 &  403 & 111 &  97 \\
28 Feb &  888 &  531 & 151 & 145 \\
29 Feb & 1128 &  615 & 191 & 217 \\
01 Mar & 1694 &  984 & 263 & 285 \\
02 Mar & 1835 & 1254 & 273 & 335 \\
03 Mar & 2502 & 1326 & 307 & 420 \\
04 Mar & 3089 & 1497 & 360 & 544 \\
05 Mar & 3858 & 1777 & 380 & 658 \\
06 Mar & 4636 & 2008 & 454 & 816 \\
07 Mar & 5883 & 2742 & 543 & 1010 \\
\hline
\end{tabular}

\medskip

{\tt Table A.IV.a.} Known cases of contagion in all of Italy (I) and in different regions: Lombardia (L), Veneto (V), Emilia-Romagna (ER) \cite{MinSal,PCrep}.
\bigskip

\begin{tabular}{|l||r|r|r|r|r||r||}
\hline
day & IC & SC & HI & Rec & Dead & Total \\
\hline
24 Feb &  27 &  101 &   94 &   1 &   5 &  229 \\
25 Feb &  35 &  114 &  162 &   1 &  10 &  322 \\
26 Feb &  36 &  128 &  221 &   3 &  12 &  400 \\
27 Feb &  56 &  248 &  284 &  45 &  17 &  650 \\
28 Feb &  64 &  345 &  412 &  46 &  21 &  888 \\
29 Feb & 105 &  401 &  543 &  50 &  29 & 1128 \\
01 Mar & 140 &  639 &  798 &  83 &  34 & 1694 \\
02 Mar & 166 &  742 &  927 & 149 &  52 & 1835 \\
03 Mar & 229 & 1034 & 1000 & 160 &  79 & 2502 \\
04 Mar & 295 & 1346 & 1065 & 276 & 107 & 3089 \\
05 Mar & 351 & 1790 & 1155 & 414 & 148 & 3858 \\
06 Mar & 462 & 2394 & 1060 & 523 & 197 & 4636 \\
07 Mar & 567 & 2651 & 1843 & 589 & 233 & 5883 \\
\hline
\end{tabular}

\medskip

{\tt Table A.IV.b.} Known cases of contagion in all of Italy (cumulative), according to treatment. IC: patients in Intensive Care units; SC: patients in Standard Care units; HI: infected people in home isolation; Rec: recovered \cite{MinSal,PCrep}.
\bigskip

\begin{tabular}{|l||r|r|r||r||}
\hline
day & IC + SC & Home & Total & $\rho$ \\
\hline
24 Feb &  128 &   94 &  222 & 0.42 \\
25 Feb &  149 &  162 &  311 & 0.52 \\
26 Feb &  164 &  221 &  385 & 0.57 \\
27 Feb &  304 &  284 &  588 & 0.48 \\
28 Feb &  409 &  412 &  821 & 0.50 \\
29 Feb &  506 &  543 & 1049 & 0.52 \\
01 Mar &  779 &  798 & 1577 & 0.51 \\
02 Mar &  908 &  927 & 1835 & 0.51 \\
03 Mar & 1263 & 1000 & 2263 & 0.44 \\
04 Mar & 1641 & 1065 & 2706 & 0.39 \\
05 Mar & 2141 & 1155 & 3296 & 0.35 \\
06 Mar & 2856 & 1060 & 3916 & 0.27 \\
07 Mar & 3218 & 1843 & 5061 & 0.36 \\
\hline
\end{tabular}

\medskip
{\tt Table A.IV.c.} Known cases of contagion in all of Italy; comparison of the number of hospitalized patients versus those in home isolation. The $\rho$ ratio in the last column is that of Home/(IC + SC + Home), and after fluctuating around 0.5 for various days is now decreasing. Elaboration from Table A.IV.b.
\bigskip

\begin{tabular}{|l||r|r|r||r||}
\hline
day & IC & SC & Total & $\s$ \\
\hline
24 Feb &  27 &  101 &  128 & 0.21 \\
25 Feb &  35 &  114 &  149 & 0.23 \\
26 Feb &  36 &  128 &  164 & 0.22 \\
27 Feb &  56 &  248 &  304 & 0.18 \\
28 Feb &  64 &  345 &  409 & 0.16 \\
29 Feb & 105 &  401 &  506 & 0.21 \\
01 Mar & 140 &  639 &  779 & 0.18 \\
02 Mar & 166 &  742 &  908 & 0.18 \\
03 Mar & 229 & 1034 & 1263 & 0.18 \\
04 Mar & 295 & 1346 & 1641 & 0.18 \\
05 Mar & 351 & 1790 & 2141 & 0.16 \\
06 Mar & 462 & 2394 & 2856 & 0.16 \\
07 Mar & 567 & 2651 & 3218 & 0.18 \\
\hline
\end{tabular}

\medskip

{\tt Table A.IV.d.} COVID-19 patients hospitalized in Italy, in IC and in SC units. The $\s$ in the last column gives the ratio IC/(IC+SC), and fluctuates around 0.2, in line with findings in China \cite{CDC}. Elaboration from Table A.IV.b.
\bigskip

\end{center}


\section{Italy. Local data for specific areas}

\begin{center}

{\small
\begin{tabular}{||r|r|r|r||r|r|r|r||r|r|r|r||}
\hline
    LO &  CR &  PC &  PV &  BG &  BS &  MI &  PD & VE &  TV & RN &  PU \\
\hline
    23 &  36 & 29 & 55 & 110 & 126 & 320 & 94 & 86 & 89 & 34 & 36 \\
\hline
\end{tabular}
}
\medskip

{\bf Table A.V.a.} Population (in $10^4$ units) of the analyzed Departments.
\bigskip

{\small
\begin{tabular}{|l||r|r|r|r||r|r|r|r||r|r|r|r||}
\hline
   day &  LO &  CR &  PC &  PV &  BG &  BS &  MI &  PD & VE &  TV & RN &  PU \\
\hline
27 F & 159 &  91 &  63 &  36 &  72 &  10 &  15 &  59 &  14 &  22 &   6 &   3 \\
28 F & 182 & 123 &  89 &  49 & 103 &  13 &  29 &  68 &  15 &  23 &   9 &   6 \\
29 F & 237 & 136 & 138 &  55 & 110 &  14 &  30 &  81 &  19 &  35 &  15 &  11 \\
01 M & 344 & 214 & 174 &  78 & 209 &  49 &  46 & 105 &  24 &  43 &  16 &  24 \\
02 M & 384 & 223 & 212 &  83 & 243 &  60 &  58 & 135 &  48 &  72 &  19 &  33 \\
03 M & 482 & 287 & 256 & 122 & 372 &  86 &  93 & 144 &  48 &  82 &  24 &  55 \\
04 M & 559 & 333 & 319 & 126 & 423 & 127 & 145 & 162 &  59 &  86 &  33 &  72 \\
05 M & 658 & 406 & 378 & 151 & 537 & 155 & 197 & 175 &  73 &  89 &  68 & 100 \\
06 M & 739 & 452 & 426 & 180 & 623 & 182 & 267 & 198 &  85 & 103 &  93 & 126 \\
07 M & 811 & 562 & 426 & 221 & 761 & 413 & 361 & 216 & 100 & 110 & 104 & 159 \\
\hline
\end{tabular}

}

\medskip

{\tt Table A.V.b.} Known cases of contagion in specific Departments: Lodi (LO), Cremona (CR), Piacenza (PC), Pavia (PV), Bergamo (BG), Brescia (BS), Milano (MI), Padova (PD), Venezia (VE), Treviso (TV), Rimini (RN) and Pesaro (PU).

\bigskip
\end{center}

\end{appendix}

\end{document}